\pdfoutput=1
\documentclass[10pt,aps, prb,reprint,superscriptaddress, showpacs,eqsecnum]{revtex4-1}
\usepackage[utf8x]{inputenc}
\usepackage[pdftex]{graphicx}
\usepackage{amsmath,amssymb}
\usepackage{subfigure}
\usepackage{todonotes}
\newcommand{\trace}[2]{\text{Tr}_{#1}\left[#2\right]}
\newcommand{\fig}[1]{Fig.~\ref{#1}}
\newcommand{\SEC}[1]{Sec.~\ref{#1}}

\begin{document}

\title{Full counting statistics applied to dissipative Cooper pair pumping}

\author{P. Wollfarth}
\affiliation{Institut f\"ur Theorie der Kondensierten Materie, Karlsruher Institut f\"ur Technologie, 76128 Karlsruhe, Germany}

\author{I. Kamleitner} 
\affiliation{Institut f\"ur Theorie der Kondensierten Materie, Karlsruher Institut f\"ur Technologie, 76128 Karlsruhe, Germany}

\author{A. Shnirman} 
\affiliation{Institut f\"ur Theorie der Kondensierten Materie, Karlsruher Institut f\"ur Technologie, 76128 Karlsruhe, Germany}
\affiliation{DFG-Center for Functional Nanostructures (CFN), 76128 Karlsruhe, Germany}
  
\begin{abstract}
We calculate the charge transport in a flux biased dissipative Cooper 
pair pump using the method of full counting statistics (FCS). 
This is instead of a more traditional technique of integrating a very small 
expectation value of the instantaneous current over the pumping period. 
We show that the rotating wave approximation (RWA), which fails in the traditional 
technique, produces accurate results within the FCS method.
\end{abstract}

\pacs{03.65.Vf, 03.65.Yz, 05.30.-d, 85.25.Cp}

\date{\today}

\maketitle

\section{Introduction}
\label{sec:introduction}

Mesoscopic quantum systems evolving periodically and adiabatically in time provide an opportunity to study geometric effects, e.g., the Berry phase\cite{Berry-1984,Shapere-1989}, in a dissipative environment.
In particular, normal\cite{Geerligs-1990,Pothier-1992} and superconducting\cite{Geerligs-1991} charge pumping devices belong to this class of systems. The relation between the pumped charge and the Berry phase 
was proved for superconducting pumps by Aunola et al.\cite{Aunola-2003}. In experiments with qubits
\cite{Leeks-2007,Wallraff-2012} and atoms\cite{Oberthaler-1999}, in order to observe the Berry phase, superpositions of energy eigenstates must be generated. In contrast, in the charge pumping devices 
the system remains all the time close to the ground state. What is, then, being measured is the variation 
of the ground state's Berry phase with the change of the phase bias of the system. The first experimental determination of the Berry phase in a superconducting Cooper pair pump (CPP) was performed by M\"ott\"onen \textit{et al.}\cite{Mottonen-2008}. 

The adiabatically controlled CPPs are susceptible to environmental noise and dissipation. These effects have 
been investigated\cite{Solinas-2010,Pekola-2010}. One common approach to describe the influence 
of dissipation in such systems is to employ the quantum master equation technique. This way one obtains the time evolution of the reduced density matrix of the system. However employing RWA in order to obtain a Markovian master equation in Lindblad form\cite{Breuer}, which grants complete positivity, leads to incorrect results for the pumped charge\cite{Kamleitner-2011,Salmilehto-2012}. To obtain the correct pumped charge one has to go beyond the RWA\cite{Kamleitner-2011}, i.e. include the fast
rotating terms of the interaction picture master equation.
The very simple reason for this is the fact that the tiny 
instantaneous value of the current is determined by the tiny off-diagonal elements of the density matrix 
dropped within RWA. 

In this paper we employ an alternative formalism of FCS\cite{levitov-1996}. This is achieved by including (theoretically) an ideal measuring apparatus in our description of the electrical circuit, i.e., a capacitor 
with infinite capacitance. Such a capacitor has zero impedance at all frequencies, i.e., it does not disturb the 
circuit's dynamics. Yet it allows to introduce formally the number of Cooper pairs that have passed through the system and the conjugate counting field. It turns out, that within this approach the RWA can be used safely and we obtain a master equation for the combined system of CPP and the measuring device in a Lindblad form. Thus,
we develop a formalism which results simultaneously in the correct pumped charge and a positive definite 
reduced density matrix of the system.
We use the results of Ref. \onlinecite{Kamleitner-2011} as a benchmark (obtained by going beyond the RWA) 
and obtain a perfect matching with our current results. 
In this context it should be mentioned that in superconducting phase biased devices the interpretation 
of the higher FCS cumulants is tricky\cite{Belzig-Nazarov-2001}. 

After providing a brief introduction to the CPP system in \SEC{sec:cpp}, in \SEC{sec:cpp:withdevice} we 
introduce a {\it gedanken} measuring device, which allows us to use the methods of FCS. We briefly demonstrate the power of this approach by reproducing the connection between the Berry phase and the pumped charge for a closed system\cite{Aunola-2003}. In \SEC{sec:meq} we derive a Markovian quantum master equation for the reduced density matrix of the combined system consisting of CPP and the measuring device. It should be emphasized that in the derivation the RWA was used with respect to the fast rotating terms in the interaction picture. In \SEC{sec:results} we present and discuss the obtained results and compare them with previous calculations, which have been done via integrating the current\cite{Kamleitner-2011}. In Section \ref{sec:conclusions} we conclude. 

\section{The System}
\label{sec:cpp}
\begin{figure}[h!!!!!!!!!!!!!!!!!]
\begin{center}
 \includegraphics[width=0.48\textwidth]{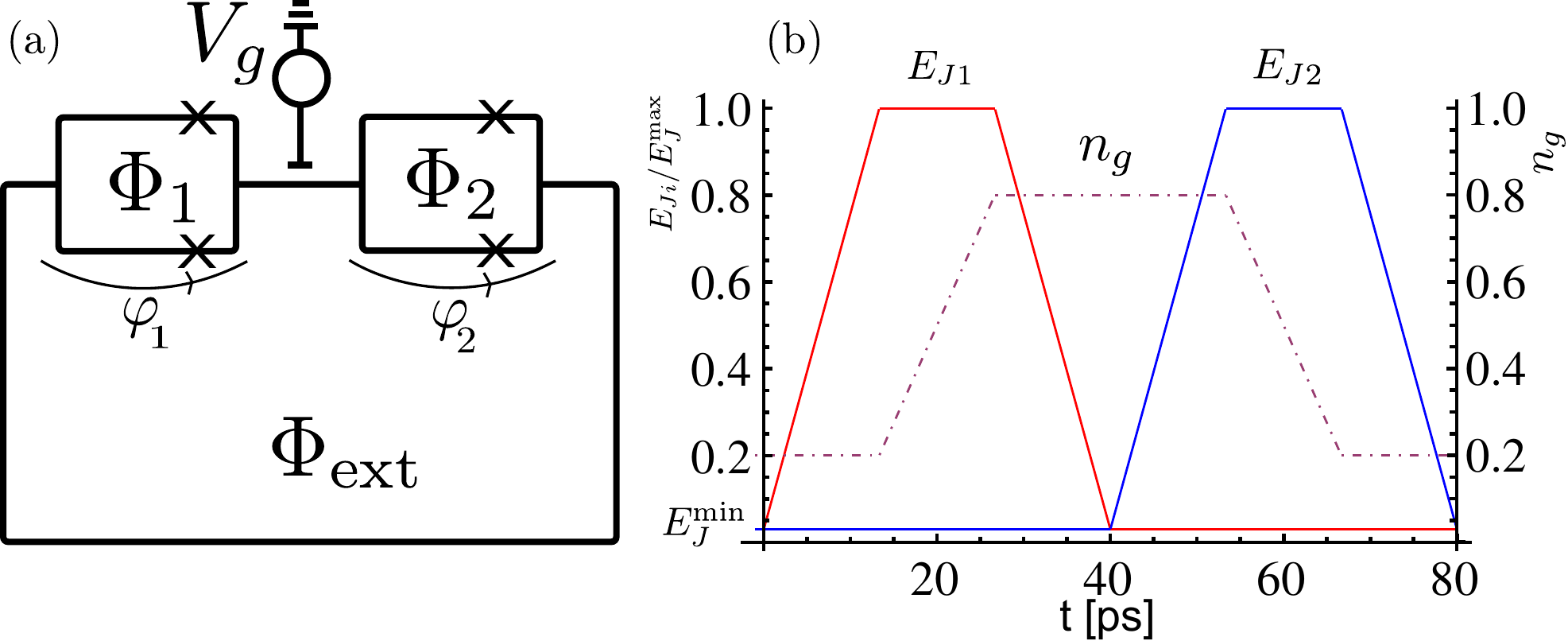}
\caption{(a) Flux biased Cooper pair pump consisting of two SQUIDs enclosing a superconducting island. The SQUIDs are controllable by external fluxes $\Phi_i$ and $\Phi_\text{ext}$. The island is driven by a gate voltage $V_g$. (b) Evolution of parameters $E_{J1}$, $E_{J2}$ and $n_g$ during a pumping cycle of period $\tau =80\text{~ps}$. }
\label{pic:circuit}
 % circuit.pdf: 0x0 pixel, 300dpi, 0.00x0.00 cm, bb=
\end{center}
\end{figure}

The system we are studying is the so called Cooper pair sluice\cite{Niskanen-2003,mottonen-2006,Mottonen-2008} depicted in \fig{pic:circuit} (a). It consists of two SQUIDs enclosing a superconducting island. 
The island is controlled via a gate voltage $V_g$ whereas the SQUIDs are manipulated by external fluxes $\Phi_i$. The system is well studied theoretically\cite{mottonen-2006} and was also realized experimentally\cite{Mottonen-2008}. The Hamiltonian is given by \cite{Niskanen-2003} 
\begin{align}
H= E_C (\hat{n}-n_g)^2 &-  E_{J,1}(\phi_1)\cos\left(\frac{\phi_\text{ext}}{2} +{\varphi}\right)\nonumber \\
&-E_{J,2}(\phi_2)\cos\left(\frac{\phi_\text{ext}}{2} -{\varphi}\right).
\end{align}

Here, $\hat{n}$ is the operator of number of excess Cooper pairs on the island, $n_g=C_g V_g$ is the gate charge, $E_C= (2e^2)/(C)$ denotes the charging energy with $C$ being the capacitance of the island, $E_{J,i}(\phi_i)$ are the tunable Josephson energies of the SQUIDs, $\phi_\text{ext}$ is the total phase of the circuit and 
${\varphi}=\left(\varphi_1 -\varphi_2\right)/2$ corresponds to the phase on the superconducting island. The connection between the fluxes and phases is given by $\phi_i = 2\pi\Phi_i/\Phi_0$ with $\Phi_0$ being the magnetic flux quantum. Since the phase ${\varphi}$ and the number operator $\hat{n}$ are conjugated variables, the commutation relation $\left[\hat{n},e^{i\varphi}\right] = e^{i\varphi}$ holds.

To pump Cooper pairs through the device, the external parameters $E_{J,i}$ and $n_g$ have to be altered cyclically and adiabatically. An example of a pumping cycle is given in \fig{pic:circuit} (b).  In the beginning of the cycle the SQUIDs are closed and there is no gate voltage applied to the island. First, the left SQUID is opened by increasing its Josephson energy to the value of $E_J^\text{max}$. After the left SQUID is opened, the gate charge $n_g$ is increased to pull a Cooper pair onto the island. In the next step the left SQUID is shut off and the right SQUID is opened. Decreasing the gate charge finally pushes the Cooper pair into the right lead, so that altogether one Cooper Pair is pumped through the device. 

Real SQUIDs are not perfectly closable, therefore we consider a residual Josephson energy $E_J^\text{min}=0.03 E_J^\text{max}$, which is a realistic experimental value. This causes a supercurrent flowing through the island in addition to the pumped current.

\begin{figure}[h!!!!!!!!!!!!!!!!!]
\begin{center}
 \includegraphics[width=0.3\textwidth]{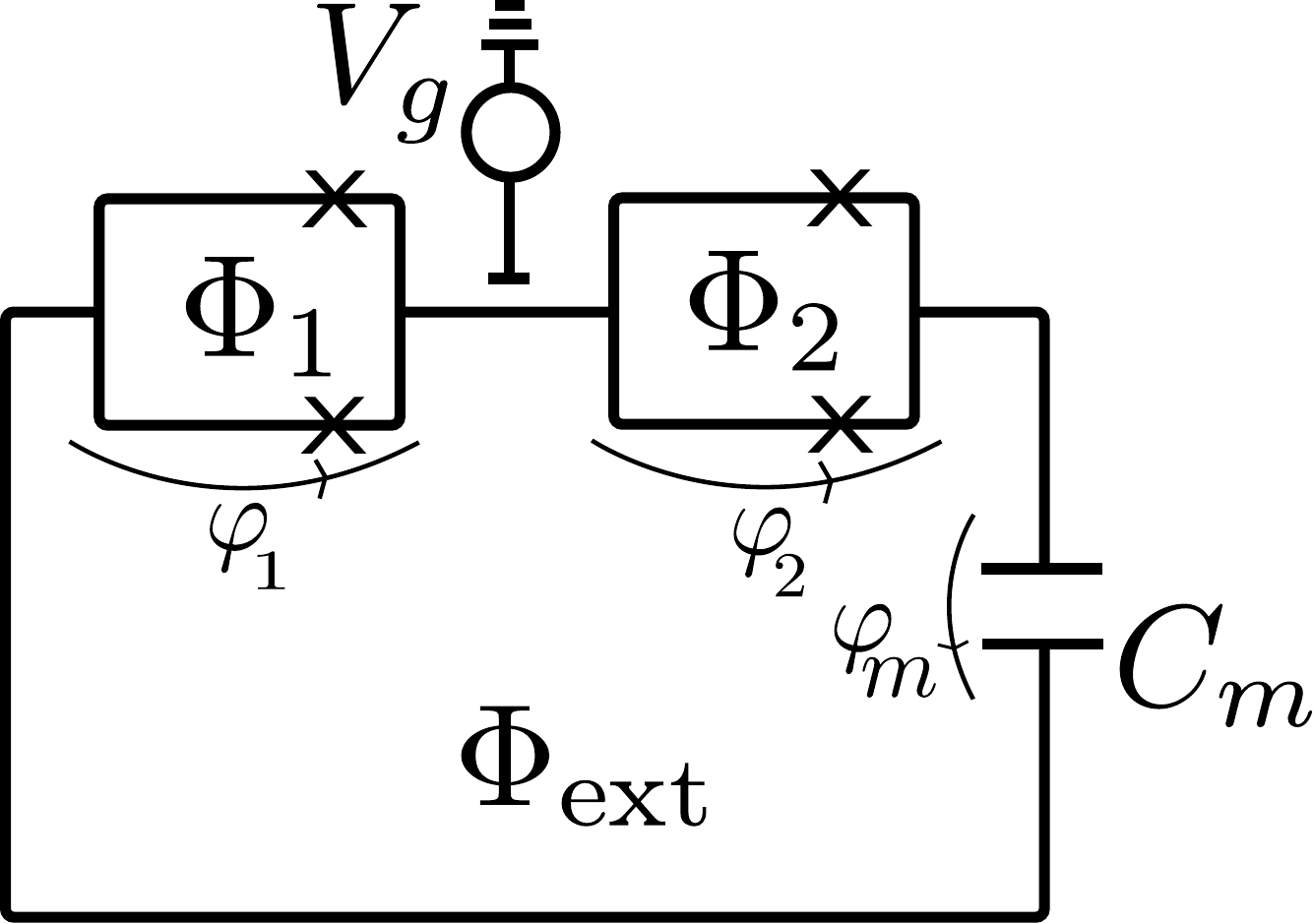}
\caption{Cooper pair pump as in \fig{pic:circuit}. An infinite capacitor $C_m$ has been implemented as a measuring device for transferred charge creating a new phase $\varphi_m$ to the system.}
\label{pic:circuit:withdevice}
 % circuit.pdf: 0x0 pixel, 300dpi, 0.00x0.00 cm, bb=
\end{center}
\end{figure}  

\section{Implementation of the measuring device}
\label{sec:cpp:withdevice}
To calculate the transferred charge, we use the technique of full counting statistics\cite{levitov-1996} (FCS). 
In particular, we consider a {\it gedanken} experiment in which we add a charge measuring device to the circuit. This device is a capacitor $C_m$ described by a phase variable $\varphi_m$ (see \fig{pic:circuit:withdevice}). To minimize the disturbance of the circuit by the measuring device, we choose $C_m \rightarrow \infty$, such that no charging energy term is added to $H$. 

To demonstrate the functionality of the capacitor as a measuring device, we have a closer look at the (reduced) density matrix $\rho_M$ of the measuring device. Whenever $N$ Cooper pairs reach the capacitor, its state changes as $|\varphi_m\rangle \rightarrow e^{iN\varphi_m}|\varphi_m\rangle$. Hence the off-diagonal element $|\varphi_m\rangle \langle \varphi_m'|$ of the density matrix changes as
\begin{align}
 |\varphi_m\rangle \langle \varphi_m'| \xrightarrow{\text{N Cooper pairs transported}}e^{iN(\varphi_m -\varphi_m')}|\varphi_m\rangle \langle \varphi_m'|.
\end{align}
If we now assume a probability distribution $P_N$ of $N$ transferred charges, the corresponding transformation of the density matrix of the measuring device reads
\begin{align}
  |\varphi_m\rangle \langle \varphi_m'| \rightarrow \underbrace{\sum_N P_N e^{iN(\varphi_m -\varphi_m')}}_{\chi(\lambda)}|\varphi_m\rangle \langle \varphi_m'|,
 \end{align}
where $\chi(\lambda)$ is the desired cumulant generating function (CGF) to calculate the transferred charge. For the counting field we get $\lambda =\varphi_m -\varphi_m'$. In analogy to Ref. \onlinecite{levitov-1996}, the CGF is obtained as
\begin{align}\label{eq:cgf}
 \chi(\lambda)  = \frac{\trace{P}{\langle \varphi_m |\rho_P(t) |\varphi_m'\rangle}}{\trace{P}{\langle \varphi_m |\rho_P(0) |\varphi_m'\rangle}}.
\end{align}
With the CGF, the cumulants of the number of Cooper pairs transferred onto the capacitor $C_m$ are calculated as
\begin{align}\label{eq:cumulant}
\mathcal{C}_n =(1/i)^n \partial_\lambda^n \ln(\chi(\lambda))|_{\lambda=0} .
\end{align}
In particular, we are interested in the charge $Q = 2e \,\mathcal{C}_1$. After the incorporation of the capacitor $C_m$ into the system, the new Hamiltonian for the system plus measuring device reads

\begin{align}
 H_{PM} = \int d\varphi_m H_P(\varphi_m) \otimes |\varphi_m \rangle \langle \varphi_m|,
\end{align}
with 
\begin{align}
  H_P(\varphi_m)= E_C (\hat{n}-n_g)^2 & - E_{J,1}(\phi_1)\cos\left(\frac{\phi_\text{ext}}{2} +\varphi- \frac{\varphi_m}{2}\right)\nonumber \\
&-E_{J,2}(\phi_2)\cos\left(\frac{\phi_\text{ext}}{2} -\varphi- \frac{\varphi_m}{2}\right).
\end{align}
The property of $H_{PM}$ being diagonal in $\varphi_m$, i.e., the absence of the conjugate charge is due to $C_m \rightarrow \infty$. This means that $\varphi_m$ is a constant of motion. 

In the limit of $E_C \gg E_J^\text{max}$ and $n_g \in [0,1]$, it is possible to restrict the Hamiltonian to a two level Hilbert space of two charge states\cite{mottonen-2006} $\{|0\rangle, |1\rangle\}$, where $|0\rangle (|1\rangle)$ correspond to zero (one) excess Cooper pair on the island. The two-level Hamiltonian reads
\begin{align}
 H_P^{2ls}&=-1/2 \Bigg[E_C(1-2n_g)\sigma_z \nonumber \\
&+ (E_{J1}+E_{J2})\cos\left(\frac{\phi_\text{ext}-\varphi_m}{2}\right) \sigma_x \nonumber\\
&+ (E_{J1}-E_{J2})\sin\left(\frac{\phi_\text{ext}-\varphi_m}{2}\right)\sigma_y\Bigg].
\end{align}
By diagonalizing this Hamiltonian we get the energies $E_{g/e}$ for the corresponding ground- and excited states 
\begin{align}\label{eq:eigenzust}
|g(\varphi_m)\rangle &= ae^{i\gamma}|0\rangle + b |1\rangle, \\
|e(\varphi_m)\rangle &= be^{i\gamma}|0\rangle - a |1\rangle.
 \end{align}
Here, the amplitudes $a$ and $b$ as well as the phase $\gamma$ are real numbers satisfying $a^2 +b^2 = 1$, 
which depend on the controlled parameters of the Hamiltonian ($E_{J1}, E_{J2},n_g$) as well as on the ``measuring" phase $\varphi_m$. They are explicitly given by
\begin{align}
 a^2 &= 1-b^2  =\frac{1}{2} \left(1- \frac{\left(n_g-1/2\right)}{\sqrt{\left(n_g-1/2\right)^2 + \frac{E_{12}^2}{4 E_C^2}}}\right), \\
\gamma &= \arctan\left[\frac{E_{J2} -E_{J1}}{E_{J1}+E_{J2}}\tan\left(\frac{\phi_\text{ext}-\varphi_m}{2}\right)\right],
\end{align}
where we introduced the abbreviation $E_{12} = \sqrt{E_{J1}^2 + E_{J2}^2 +2E_{J1}E_{J2} \cos\left(\phi_\text{ext}-\varphi_m\right)}$. A similar representation without the measuring device was used in Ref. \onlinecite{mottonen-2006}.

As a first verification of the technique of FCS, we calculate the transferred charge in the closed system, i.e. without coupling to an environment. We consider one pumping cycle of period $\tau$. It follows from eq. \eqref{eq:cgf}, that the CGF is given by 
\begin{align}\label{eq:cgftotal}
 \chi(\lambda) = e^{-i\Theta_{\text{D}}(\varphi_m)}e^{-i\Theta_{\text{B}}(\varphi_m)}e^{i\Theta_{\text{B}}(\varphi_m')}e^{i\Theta_{\text{D}}(\varphi_m')},
\end{align}
where we assume an adiabatic ground state evolution, i.e. $|\Psi(\tau,\varphi_m) \rangle \approx e^{-i\Theta_{\text{D}}(\varphi_m)}e^{-i\Theta_{\text{B}}(\varphi_m)}|g(0,\varphi_m)\rangle$. Here  $\Theta_\text{D} = \int_0^\tau dt E_g(\varphi_m,t)$ is the dynamical phase of the ground state of the system and $\Theta_\text{B}=-\oint \langle g(\vec{q})|\nabla_{\vec{q}}|g(\vec{q})\rangle d \vec{q}$ is the geometric Berry phase of the ground state\cite{Berry-1984}. The vector $\vec{q}$ belongs to the reduced parameter space spanned by two 
parameters $\vec{q}=\left(a^2(\varphi_m),\gamma(\varphi_m)\right)$. From the first cumulant we find for the transferred charge
\begin{align}\label{eq:bare:charge:tot}
 Q_\text{tot} = -2e\, \mathcal{C}_1 = 2e \left(\partial_{\varphi_m}\Theta_\text{D} +\partial_{\varphi_m} \Theta_\text{B}\right).
\end{align}
We identify the first term of eq. \eqref{eq:bare:charge:tot} as the charge $Q_\text{S}=2e\partial_{\varphi_m}\Theta_\text{D}$ due to the super-current. Because the pumped charge solely depends on the time dependence of the system, the second term of eq. \eqref{eq:bare:charge:tot} $Q_\text{P}=2e\partial_{\varphi_m} \Theta_\text{B}$ corresponds to the pumped charge. This concurs with the previous results\cite{Aunola-2003}.

\section{Lindblad master equation}
\label{sec:meq}
In this section, we derive a Markovian master equation of the Lindblad form to describe the influence of dissipation on our system. The derivation is performed along the lines of Ref. \onlinecite{Kamleitner-2011}, a detailed calculation including the RWA can be found in appendix \ref{sec:appendix:meq}. The main source of dissipation are fluctuations $\delta V_g$ of the gate voltage $V_g=V_g^0 +\delta V_g$, which results in a coupling proportional to $\sigma_z$ in the charge basis. These are, e.g., due to the dissipative parts of the controlling circuit impedance. 
Also material dielectric losses can be effectively modeled this way.
In previous calculations\cite{Kamleitner-2011} the secular approximation (RWA), well known from the field of quantum optics, turned out to be inadequate (non charge conserving)\cite{Salmilehto-2012}. 

Our starting point is a Markovian master equation in the interaction picture\cite{Breuer}
\begin{align}\label{eq:meqstart}
  \frac{d}{dt} \rho_{PM}(t) = \!-\!\!\int_0^\infty  \!\!\!\!\!\!ds\, \trace{B\!\!}{H_I(t),\left[H_I(t\!-\!s),\rho_{PM}(t)\!\otimes\!\rho_B\right]}.
\end{align}
Here, $\rho_{PM}(t)$ is the density matrix of the combined system of pump (P) plus measuring device (M) and $H_I(t)=A_{PM}(t) \otimes B(t)$ describes the interaction between the system, represented by $A_{PM}(t)$and the bath, represented by $B(t)$. The explicit form of $A_{PM}$ in the Schr\"odinger picture is
\begin{align}
 A_{PM} = \int d\varphi_m  E_C \sigma_z \otimes |\varphi_m\rangle\langle \varphi_m|.
\end{align}
 The density matrix $\rho_B$ of the bath is assumed to correspond to a thermal equilibrium state.

To perform the RWA we need to consider eq. \eqref{eq:meqstart} in an explicit basis. Because of the cyclic time dependence of the control parameters with period $\tau$, we use the Floquet states $|\phi_n(t)\rangle$ for this purpose. These states are cyclic solutions (up to a phase) of the Schr\"odinger equation with the property $|\phi_n(t+\tau)\rangle \propto |\phi_n(t)\rangle$. They can be obtained numerically and have the useful property\cite{Kamleitner-2011}, that they implicitly contain super adiabatic corrections to the instantaneous eigenstates\cite{Berry-1987}. For a review on Floquet theory, see Ref. \onlinecite{Grifoni-1998}.

In order to compute the CGF according to eq. \eqref{eq:cgf} it is necessary to evaluate the time evolution of the reduced density matrix of the Cooper pair pump including the off-diagonal elements with respect to the measuring (counting) phases $\langle \varphi_m |\rho_{PM}(t)|\varphi_m'\rangle = \rho_P^{\varphi_m \varphi_m'}(t)$. In the limit $C_m \rightarrow \infty$, the counting fields $\varphi_m$ and $\varphi_m'$ are constants of motions. This means that the time evolution
factorizes, i.e. $\rho_P^{\varphi_m \varphi_m'}(t)$ depends only on matrix elements with the same $\varphi_m$ and $\varphi_m'$.
This significantly simplifies the analysis of the master equation as we only have to deal with $2\times2$ matrices. We obtain

\begin{widetext}
 \begin{align}%\label{eq:meq:dissi2}
  \dot{\rho}_P^{\varphi_m \varphi_m'}  &= -i  \Big[H_P^{2ls}+ H_{LS}\Big]\rho_P^{\varphi_m \varphi_m'}+i \rho_P^{\varphi_m \varphi_m'} \Big[{H'}_P^{2ls} + {H'}_{LS}\Big]
+ \gamma(0)\Big[L_0\rho_P^{\varphi_m \varphi_m'}{L_0'}^\dagger-\frac{1}{2}\left( L_0^\dagger L_0\rho_P^{\varphi_m \varphi_m'} +\rho_P^{\varphi_m \varphi_m'}{L_0'}^\dagger L_0' \right) \Big]\nonumber \\
&+  \sum_{j \neq k \in \{e,g\}}\Bigg[ \frac{\gamma(\omega_{j k}) +\gamma(\omega_{j k}')}{2} L_{j k}\rho_P^{\varphi_m \varphi_m'}{L_{j k}'}^\dagger+i\frac{\xi(\omega_{j k}) -\xi(\omega_{j k}')}{2} L_{j k}\rho_P^{\varphi_m \varphi_m'}{L_{j k}'}^\dagger \nonumber \\
&- \frac{1}{2}\Big[\gamma(\omega_{j k}) L_{j k}^\dagger  L_{j k}\rho_P^{\varphi_m \varphi_m'} + \gamma(\omega_{j k}') \rho_P^{\varphi_m \varphi_m'}{L_{j k}'}^\dagger L_{j k}' \Big] \Bigg].
\end{align}

\end{widetext}
Here, $\gamma(\omega)$ and $\xi(\omega)$ are the real- and imaginary parts of the half-sided Fourier (Laplace) transform of the bath correlation function $\gamma(\omega)/2 + i \xi(\omega) = \int_0^\infty \langle B(s) B(0) \rangle \,e^{i\omega s}\, ds$.
The transition frequencies are given by 
$\omega_{j k}\approx E_k(t)-E_j(t)$ [the explicit form is given in eq. \eqref{eq:transfreq}], 
and we introduced the Lindblad operators
\begin{align}
 L_0 (\varphi_m, t) &= E_C\sum_n \langle \phi_n|\sigma_z|\phi_n \rangle |\phi_n\rangle \langle \phi_n|, \\
 L_{jk}(\varphi_m, t) &= E_C\langle \phi_j|\sigma_z|\phi_k\rangle |\phi_j\rangle \langle \phi_k|,
\end{align}
as well as the Lamb shift 
\begin{align}
 H_{LS}(\varphi_m, t) &= E_C^2\sum\limits_{j,k} \xi(\omega_{jk}) |\langle \phi_j|\sigma_z|\phi_k \rangle |^2 \times |\phi_j\rangle \langle \phi_j|.
\end{align}
Quantities labeled with and without prime depend on $\varphi_m'$ and $\varphi_m$ respectively. The Lindblad operator $L_0 (\varphi_m, t)$ describes the \emph{pure} dephasing, whereas $L_{jk} (\varphi_m, t)$ cause relaxation and excitation. The coherent part of the master equation consists of a ``commutator" between $\rho_P^{\varphi_m \varphi_m'}$ and $H_P^{2ls}+ H_{LS}$ (this is not exactly a commutator since the counting 
phase in the Hamiltonian $H_P^{2ls}+ H_{LS}$ is different depending on which side of the density matrix 
it is placed). 

It should be emphasized that in, contrast to Ref. \onlinecite{Pekola-2010,Kamleitner-2011}, we did perform the secular approximation (RWA) concerning fast rotating terms in the interaction picture. The resulting master equation for $\varphi_m=\varphi_m'$ is therefore a Lindblad master equation. For $\varphi_m \neq \varphi_m'$ the matrix $\rho_P^{\varphi_m \varphi_m'}(t)$ has neither trace of unity nor is it hermitian, therefore, a Lindblad form cannot be expected for the off-diagonal entries.

%Finally it should be mentioned that the Floquet states used to describe the time dependency of the system are implicitly taking super adiabatic corrections into account.

\section{Results}
\label{sec:results}

\begin{figure}[h!!!!!!!!!!!!!!!!!]
\begin{center}
 \includegraphics[width=0.45\textwidth]{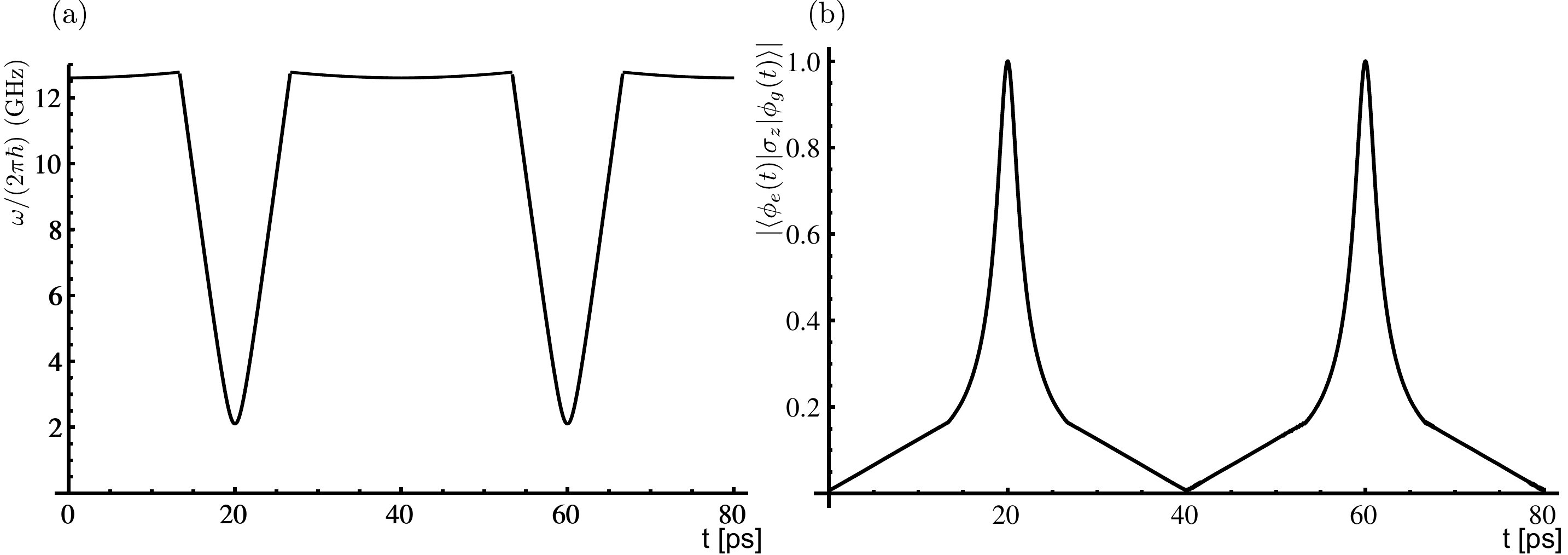}
\caption{(a) Energy splitting of the two level approximation of the system for one pumping cycle. (b) The quantity $|\langle \phi_e(t)|\sigma_z|\phi_g(t)\rangle|$ is plotted for one pumping cycle. It is proportional to the dissipation ratio and reaches its maximum at times where the energy splitting reaches its minimum. }
\label{pic:energysplitunddissi}
 % circuit.pdf: 0x0 pixel, 300dpi, 0.00x0.00 cm, bb=
\end{center}
\end{figure}

A typical description of the gate voltage fluctuation is provided by an ohmic bath, where 
the spectral density is linear in frequency. Furthermore, we assume the bath to be in thermal equilibrium $\gamma(\omega) = \gamma_0\omega\,\left(1-\exp\left(-\frac{\hbar \omega}{k_B T} \right)\right)^{-1}$. For comparison with previous work\cite{Kamleitner-2011} the duration of a pumping cycle is assumed to be $\tau = 80\text{~ps}$, the charging energy of the island is $E_C/(2\pi \hbar) = 21 \text{GHz}$ with the ratio between the maximal Josephson energy of the SQUID and the charging energy  given by $E_J^\text{max}/E_C=0.1$. The external phase is 
chosen to be $\phi_\text{ext}=-\pi/2$. In this parameter regime the two level description as well as 
adiabaticity are well justified. The energy splitting of the system along the pumping contour is shown in \fig{pic:energysplitunddissi} (a). Temperatures of order of the minimal energysplitting  $\omega_\text{min} \approx 2 \text{~GHz} \times 2\pi\hbar$ should be sufficient to significantly influence the system.

We now go on to numerically solve the master equation. In order compare  
with experiments, where multiple pumping cycles are performed, we allow the system to reach 
the quasistationary state\cite{Kamleitner-2011}. To do so, we start pumping in the ground state. After a certain number $N$ of pumping cycles ($N$ depends on temperature and the coupling strength to the dissipative bath) 
the system approaches a quasi-stationary state in which the density matrix is periodic, 
$\rho_P(t+\tau)=\rho_P(t)$. We then investigate the charge transferred during time $t$ after the quasi-stationary 
state is reached. It is obtained from equations \eqref{eq:cgf} and \eqref{eq:cumulant} and is given by
\begin{align}
 Q(t)_\text{tot} =2e \frac{\partial}{\partial \lambda}\frac{1}{i} \ln\left(\frac{\trace{P}{\rho_P^{\varphi_m \varphi_m'} ( N\tau + t)}}{\trace{P}{\rho_P^{\varphi_m \varphi_m'}(N\tau)}}\right)\Bigg|_{\lambda=0},
\end{align}
where we numerically approximate the differentiation with respect to $\lambda$ by a finite difference quotient. 

\begin{figure}[h!!!!!!!!!!!!!!!!!]
\begin{center}
\includegraphics[width=\linewidth]{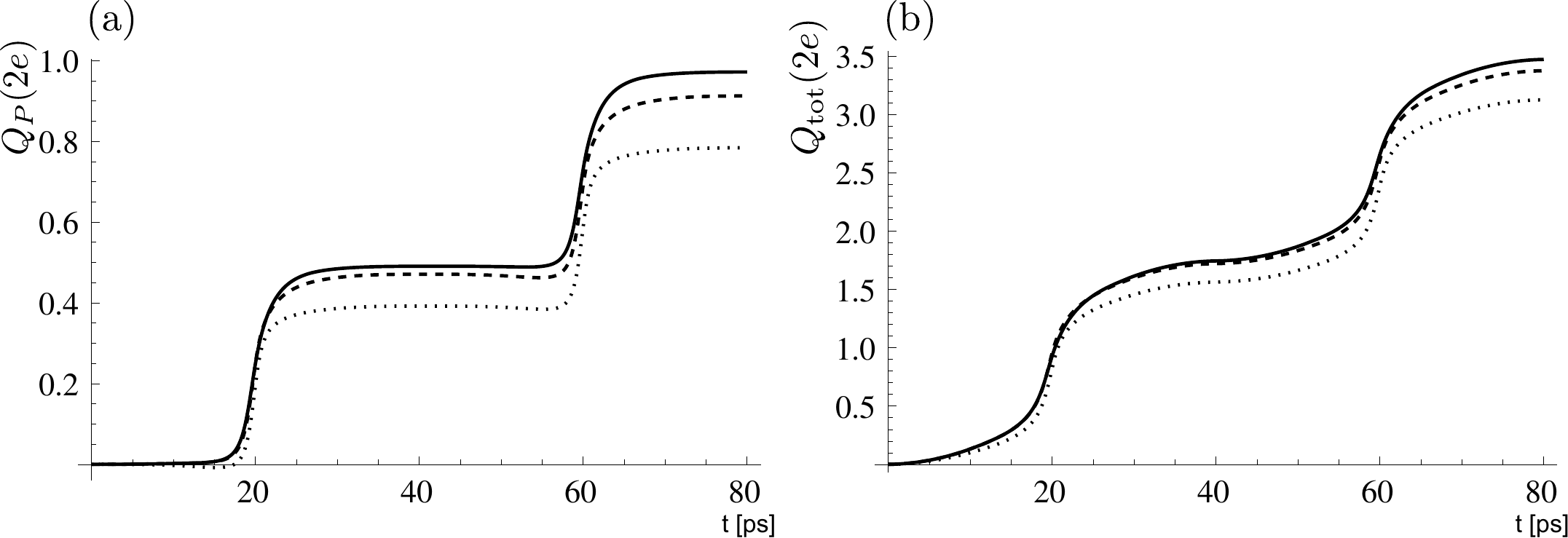}
\caption{(a) Pumped charge $Q_P$ through the island during one pumping cycle for fixed temperature $k_B T/(2\pi\hbar) = 2\text{~GHz}$ and for different coupling strengths $E_C^2 \gamma_0= 0.2$(solid), $E_C^2 \gamma_0= 0.1$(dashed) and $E_C^2 \gamma_0= 0.05$(dotted) plotted over time. As expected, charge is only pumped when the gate voltage is changed. The weaker the coupling to the environment, the more the pumped charge tends to decrease. (b) The total charge $Q_\text{tot}$ transferred through the island. The temperature is $k_B T/(2\pi\hbar) = 2\text{~GHz}$ and it is plotted for different coupling strengths $E_C^2 \gamma_0= 0.2$(solid), $E_C^2 \gamma_0= 0.1$(dashed) and $E_C^2 \gamma_0= 0.05$(dotted). The total charge also tends to decrease as the coupling strength is decreased. }
\label{pic:qpumptotcoup}
 % circuit.pdf: 0x0 pixel, 300dpi, 0.00x0.00 cm, bb=
\end{center}
\end{figure}

The pumped charge can be distinguished from the supercurrent by the fact that the former changes sign when the pumping cycle is traversed in the opposite direction, while the latter does not. The pumped charge was calculated as $Q_P = (Q_\text{tot}-\bar{Q}_\text{tot})/2$, where $\bar{Q}_\text{tot}$ is the charge transferred when the pumping cycle is traversed in the opposite direction. It is plotted as a function of time within one pumping cycle in \fig{pic:qpumptotcoup} (a). 

To understand the behavior of the pumped charge as a function of the coupling strength $E_C^2 \gamma_0$ 
to the gate voltage fluctuations we analyze energy splitting during the pumping 
cycle (\fig{pic:energysplitunddissi} (a) ) and the matrix element $|\langle \phi_e(t)|\sigma_z|\phi_g(t)\rangle|$ (\fig{pic:energysplitunddissi} (b)). In a time independent situation the relaxation rate would be proportional to $|\langle \phi_e(t)|\sigma_z|\phi_g(t)\rangle|^2$. Upon inspection of \fig{pic:energysplitunddissi} (a) and (b) it is obvious that the influence of dissipation reaches a maximum, whenever the energy splitting is minimal. On the other hand, whenever the energy splitting is large, the relaxation rate is low and the system has not enough time to relax. This indicates that a stronger coupling to the environment should support ground state pumping \cite{Pekola-2010}.
We note that the pumped current is largest at times when the voltage parameter $n_g$ crosses the charging energy degeneracy point at $n_g = 0.5$. The reason is that at these times the state of the system changes at the highest rate. Furthermore, we see that decreasing the coupling to the environment leads to a smaller pumped charge. This can be explained by the fact that with a low coupling to the environment the system gets easily excited during the times with low energy splitting. At times with high energy splitting, the relaxation rate becomes too small for the excitation to relax. After several cycles the quasi-stationary population of the excited state can therefore be quite large. As the excited state carries charge in the opposite direction the total pumped charge is reduced.

A similar trend is also visible in the total charge transferred 
through the island as depicted in \fig{pic:qpumptotcoup} (b). We see that a supercurrent flows through the island at all times. It reaches its maximum at $t=20, 60\text{ps}$, where the gate voltage is $n_g=0.5$. Comparing both figures in detail we observe that the transferred charge due to the super current is also influenced by the dissipative environment. This is apparent, as the decrease of the pumped charge is smaller than the decrease of the total transferred charge. 

\begin{figure}[h!!!!!!!!!!!!!!!!!]
\begin{center}
 \includegraphics[width=\linewidth]{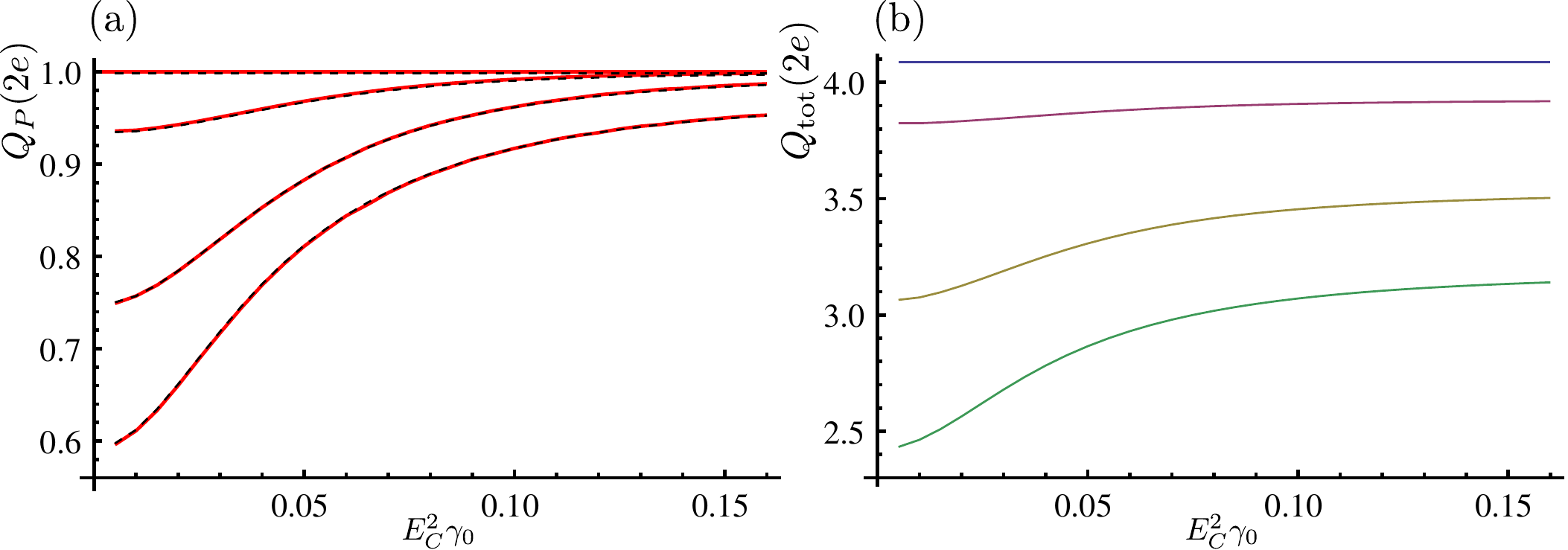}
\caption{(a): Pumped charge $Q_P$ through the island as a function of coupling strength $E_C^2 \gamma_0$. The charge is plotted for different temperatures $k_BT/(2\pi\hbar)= 0,1,2,3 \text{~GHz}$ (from top to bottom). The solid red lines represent calculations for perfectly closable SQUIDs, i.e. $E_J^\text{min}=0$, whereas the dashed black lines were made for realistic SQUIDs. The weaker the coupling to the environment, the more the pumped charge tends to decrease. In Fig. (b) the total charge is depicted as a function of the coupling strength for different temperatures $k_BT/(2\pi\hbar)= 0,1,2,3 \text{~GHz}$ (from top to bottom). }
\label{pic:qovercoup}
 % circuit.pdf: 0x0 pixel, 300dpi, 0.00x0.00 cm, bb=
\end{center}
\end{figure}

In \fig{pic:qovercoup} (a) the pumped charge $Q_P(\tau)$ for one complete pumping cycle is plotted as a function of the coupling strength for different temperatures. As mentioned before, for weaker coupling strengths as well as for higher temperatures the pumped charge tends to decrease. As was expected\cite{Pekola-2010}, a zero 
temperature environment stabilizes the adiabatic theorem, i.e., does not influence the pumped charge.

The solid red lines in \fig{pic:qovercoup} (a) represent the calculation with perfectly closable SQUIDs, 
i.e., $E_J^\text{min}=0$, whereas the dotted black lines where calculated for realistic SQUIDs with $E_J^\text{min}/E_J^\text{max}=0.03$. Since both curves almost perfectly match, the pumped charge is hardly influenced by the residual Josephson energy $E_J^\text{min}$ of the SQUID. Increasing the coupling to the environment supports the charge transfer which is in agreement with the conclusions of Ref. \onlinecite{Pekola-2010}. \fig{pic:qovercoup} (b) depicts the total charge $Q_\text{tot}(\tau)$ transferred in one complete pumping cycle. As was expected, even for zero temperature environment the total charge is not quantized. It shows a similar behavior as the pumped charge when varying in temperature and coupling strength. Comparing both figures, the charge transferred due to super current only is also affected by varying temperature or coupling strength. 
 The decrease of transferred charge for higher temperatures is due to a higher population of the excited state, which is responsible for a back flowing current.
Conclusively it should be mentioned that the obtained results are in a very good agreement with the results in Ref. \onlinecite{Kamleitner-2011}\cite{footnote}. However, in \onlinecite{Kamleitner-2011} these results could only be obtained by including the non-secular terms, while our novel method using FCS allows for the use of the RWA.

\section{Conclusions}
\label{sec:conclusions}
We explored an alternative way to calculate the amount of charge transferred in Cooper pair pumps. 
It turned out that introducing the counting fields allows us to perform the usual RWA and use the master 
equation of the Lindblad form. We conjecture that FCS is in general the instrument of choice for charge calculation whereas charge calculations via the current operator require a much more careful treatment. 
Our results are obtained numerically at this stage. A general analytic argument for why FCS in combination with RWA produces accurate results, will be the subject of future work.

\acknowledgements{This work was funded by the EU FP7 GEOMDISS project.}

% Since the eigenstates $|n(t)\rangle$ of the system are periodic in time, the following approximations hold:
% \begin{subequations}
% \begin{align}
%  |\phi_n (t)\rangle &= |n (t)\rangle + \mathcal{O}(\mathcal{A}),\\
% \bar{E}_n(t) &= E_n(t) + \mathcal{O}(\mathcal{A}^2),\\
% \bar{\theta}_B &= \theta_B +\mathcal{O}(\mathcal{A}).
% \end{align}
% \end{subequations}
% Here, $\mathcal{A}$ is the adiabatic parameter, 

%%%%%%%%%%%%%%%%%%%%%%%%%%%%%%%%%%%%%%%%%%%%%%%%%%%%%%%%%%%%%%%%%%%%%%%%%
%%% Appendices appear beneath this line

\appendix

\section{Derivation of the Master equation}
\label{sec:appendix:meq}
As the system evolves periodic in time, we use the basis of modified adiabatic Floquet states\cite{Kamleitner-2011} $|\phi_n(t)\rangle$. For a strictly adiabatic evolution, the system stays in one of its adiabatic eigenstates $|n(t_0)\rangle$. The modified Floquet states originate from the adiabatic eigenstates but include non-adiabatic corrections (for details see\cite{Kamleitner-2011}) . 

Our starting point is a Markovian master equation in the interaction picture\cite{Breuer}
\begin{align}\label{eq:meq:markov}
 \frac{d}{dt} \rho_{PM}(t) = \!-\!\!\int_0^\infty  \!\!\!\!\!\!ds\, \trace{B\!\!}{H_I(t),\left[H_I(t\!-\!s),\rho_{PM}(t)\!\otimes\!\rho_B\right]}.
\end{align}
To describe the time evolution of the system, we introduce the time evolution operator $U = \sum_n |\Psi_n(t)\rangle \langle \phi_n(0)|$, where $|\Psi_n(t)\rangle$ is the time dependent state of the system. While using the adiabatic Floquet states, we rewrite the time evolution operator as
\begin{align}\label{eq:time_evolution_operator}
 U = \sum_n e^{-i\int_0^t dt' E_n(t')}e^{-i\Theta_B^nt/\tau}|\phi_n(t)\rangle,
\end{align}
where $E_n(t)$ are the instantaneous energies of the system and $\Theta_B^n$ is the geometric phase. With the help of eq. \eqref{eq:time_evolution_operator} we are able to perform a Fourier series expansion for $\langle\phi_n(t)|A_P|\phi_m(t)\rangle$ as
\begin{align}
 A_P(t) =\sum_{n,n',k} e^{i\Omega kt} A_{nn',k}|\phi_n\rangle\langle\phi_n'| e^{-i\int_0^t dt'\omega_{nn'}(t')},
\end{align}
with $\Omega =2\pi/\tau$. The Fourier coefficients are given by
\begin{align}
 A_{nn',k}= \frac{1}{\tau} \int_0^\tau dt e^{-i\Omega kt} \langle \phi_n(t)|A_{P} |\phi_{n'}(t)\rangle.
\end{align}
and the transition frequencies read
\begin{align}\label{eq:transfreq}
  \omega_{nn'}(t) = E_{n'}(t)-E_{n}(t) +(\Theta_B^{n'}-\Theta_B^n)/\tau.
\end{align}
Eq. \eqref{eq:meq:markov}, thus, becomes
\begin{widetext}
 \begin{align}
 \frac{d}{dt}\rho_{PM}(t)\!&=\!\sum_{nn',jj',k,l} \Bigg[\Gamma (\omega_{jj'}\!+\!\Omega k)A_{nn',k}(\varphi_m)A_{jj',l}(\varphi_m') e^{i\int_0^t dt'(\omega_{jj'}(t')-\omega_{nn'}(t') -\Omega(k+l))} \nonumber\\
&\times \bigg(|\phi_{n}\rangle \langle \phi_{n'}|\rho_{PM}(t) |\phi_{j'} \rangle \langle \phi_{j}| - |\phi_{n}\rangle\langle \phi_{n'}|\phi_{j'}\rangle \langle \phi_j|\rho_{PM}(t)\bigg)\nonumber \\
&+ \Gamma^*(\omega_{jj'}+\Omega k) A^*_{nn',k}(\varphi_m)A^*_{jj',l}(\varphi_m')e^{-i\int_0^t dt'(\omega_{jj'}(t')-\omega_{nn'}(t')-\Omega(k+l))}\nonumber\\
&\times \bigg(|\phi_j \rangle\langle \phi_{j'}|\rho_{PM}(t)|\phi_{n'}\rangle\langle\phi_n|-\rho_{PM}(t) |\phi_j\rangle\langle\phi_{j'}|\phi_{n'}\rangle\langle\phi_n|\bigg) \Bigg],
\end{align}
\end{widetext}
with $\Gamma(\omega_{jj'} + \Omega k)$ being the half-side Fourier (Laplace) transform of the bath correlation functions. These are given explicitly by
\begin{align}\label{eq:Gamma:diss2}
 \Gamma (\omega_{m m'}+\Omega k) &=\!\int_0^\infty \!\!ds \,\langle B(s)B(0)\rangle e^{i\int_{t-s}^t dt'(\omega_{nn'}(t')+\Omega k)}\nonumber\\
&\approx \int_0^\infty \!\!ds \, \langle B(s)B(0)\rangle e^{i(\omega_{nn'}(t)+\Omega k)s}.
\end{align}
The approximation made in the second step assumes that the eigenenergies of the system do not vary drastically on timescales of the bath correlation time. For large values of $\Omega k \approx \omega_{nm}$ the Fourier coefficients vanish, as the states $|\phi_n(t)\rangle$ slowly oscillate. We arrive at
\begin{widetext}
 \begin{align}
 \dot{\rho}_P^{\varphi_m \varphi_m'} &= \sum_{n,j,k,l} \Gamma(\Omega k) e^{-i\Omega(k+l)t} \times\Big(A_{nn,k}(\varphi_m) A_{jj,l}(\varphi_m') |\phi_n(\varphi_m)\rangle\langle\phi_n(\varphi_m)| \rho_P^{\varphi_m \varphi_m'}(t) |\phi_j(\varphi_m')\rangle\langle \phi_j (\varphi_m')| \nonumber \\
&-A_{nn,k}(\varphi_m) A_{jj,l}(\varphi_m) \delta_{jn} |\phi_n(\varphi_m)\rangle\langle\phi_j(\varphi_m)|\rho_P^{\varphi_m\varphi_m'}(t) \Big)\nonumber\\
&+\sum_{n\neq j, k, l}\Gamma(\omega_{jn}(\varphi_m)+\Omega k) e^{-i\Omega(k+l)t} \times \Big(A_{jn,k}(\varphi_m)A_{nj,l}(\varphi_m') e^{i\int_0^t dt' (\omega_{jn}(\varphi_m,t)-\omega_{jn}(\varphi_m',t))} \nonumber \\
&\times|\phi_n(\varphi_m)\rangle\langle \phi_j(\varphi_m)|\rho_P^{\varphi_m\varphi_m'}(t) | \phi_j(\varphi_m')\rangle\langle\phi_n(\varphi_m')|- \delta_{jn} A_{jn,k}(\varphi_m)A_{nj,l}(\varphi_m) |\phi_j(\varphi_m)\rangle\langle\phi_n(\varphi_m)|\rho_P^{\varphi_m \varphi_m'}(t)\Big)\nonumber\\
&+\sum_{n,j,k,l} \Gamma^*(\Omega k) e^{i\Omega(k+l)t} \times\Big(A^*_{nn,k}(\varphi_m') A^*_{jj,l}(\varphi_m) |\phi_j(\varphi_m)\rangle\langle\phi_j(\varphi_m)| \rho_P^{\varphi_m \varphi_m'}(t) |\phi_n(\varphi_m')\rangle\langle \phi_n (\varphi_m')| \nonumber \\
&-A_{nn,k}^*(\varphi_m') A_{jj,l}^*(\varphi_m') \delta_{jn} \rho_P^{\varphi_m\varphi_m'}(t)|\phi_j(\varphi_m')\rangle\langle\phi_n(\varphi_m')| \Big)\nonumber\\
&+\sum_{n\neq j, k, l}\Gamma^*(\omega_{jn}(\varphi_m')+\Omega k) e^{i\Omega(k+l)t} \times \Big(A_{jn,k}(\varphi_m')A_{nj,l}(\varphi_m) e^{-i\int_0^t dt' (\omega_{jn}(\varphi_m,t)-\omega_{jn}(\varphi_m',t))} \nonumber \\
&\times|\phi_j(\varphi_m)\rangle\langle \phi_n(\varphi_m)|\rho_P^{\varphi_m\varphi_m'}(t) | \phi_n(\varphi_m')\rangle\langle\phi_j(\varphi_m')|- \delta_{jn} A_{jn,k}^*(\varphi_m')A_{nj,l}^*(\varphi_m') \rho_P^{\varphi_m \varphi_m'}(t)|\phi_n(\varphi_m')\rangle\langle\phi_j(\varphi_m')|\Big).
\end{align}
\end{widetext}
Assuming adiabatic evolution, only small enough values of $\Omega k$ contribute. Therefore the half-side Fourier transform of the bath correlation function can be approximated as $\Gamma(\omega_{jn}(\varphi_m) +\Omega k) \approx \Gamma(\omega_{jn}(\varphi_m))$. Using $\Gamma(\omega_{nm} = 1/2\gamma(\omega_{nm}) + i \xi(\omega_{nm})$ we arrive at
\begin{widetext}

 \begin{align}%\label{eq:meq:dissi2}
  \dot{\rho}_P^{\varphi_m \varphi_m'}  &= -i  \Big[H_P^{2ls}+ H_{LS}\Big]\rho_P^{\varphi_m \varphi_m'}+i \rho_P^{\varphi_m \varphi_m'} \Big[{H'}_P^{2ls} + {H'}_{LS}\Big]
+ \gamma(0)\Big[L_0\rho_P^{\varphi_m \varphi_m'}{L_0'}^\dagger-\frac{1}{2}\left( L_0^\dagger L_0\rho_P^{\varphi_m \varphi_m'} +\rho_P^{\varphi_m \varphi_m'}{L_0'}^\dagger L_0' \right) \Big]\nonumber \\
&+  \sum_{j \neq k \in \{e,g\}}\Bigg[ \frac{\gamma(\omega_{j k}) +\gamma(\omega_{j k}')}{2} L_{j k}\rho_P^{\varphi_m \varphi_m'}{L_{j k}'}^\dagger+i\frac{\xi(\omega_{j k}) -\xi(\omega_{j k}')}{2} L_{j k}\rho_P^{\varphi_m \varphi_m'}{L_{j k}'}^\dagger \nonumber \\
&- \frac{1}{2}\Big[\gamma(\omega_{j k}) L_{j k}^\dagger  L_{j k}\rho_P^{\varphi_m \varphi_m'} + \gamma(\omega_{j k}') \rho_P^{\varphi_m \varphi_m'}{L_{j k}'}^\dagger L_{j k}' \Big] \Bigg].
\end{align}

\end{widetext}

%%%%%%%%%%%%%%%%%%%%%%%%%%%%%%%%%%%


\begin{thebibliography}{99}
\bibitem{Berry-1984} M. V. Berry, Proc. R. Soc. A {\bf 392}, 45 (1984).
\bibitem{Shapere-1989} A. Shapere, F. Wilczek (eds) \textit{ Geometric Phases In Physics}, (World Scientific, Singapore, 1989).
\bibitem{Geerligs-1990} L. J. Geerligs, V. F. Anderegg, P. A. M. Holweg, J. E. Mooij, H. Pothier, D. Esteve, C. Urbina, and M. H. Devoret, Phys. Rev. Lett. {\bf 64}, 2691 (1990).
\bibitem{Pothier-1992} H. Pothier, P. Lafarge, C. Urbina, D. Eteve, M. H. Devoret, Europhys. Lett. {\bf 17} 249 (1992).
\bibitem{Geerligs-1991} L. J. Geerligs, S. M. Verbrugh, P. Hadley, J. E. Mooij, H. Pothier, P. Lafarge, C. Urbina, D. Esteve, and M. H. Devoret, Zeitschrift f\"ur Physik B {\bf 85}, 349 (1991). 
\bibitem{Aunola-2003} M. Aunola, J. J. Toppari, Phys. Rev. B {\bf 68}, 020502 (2003).
\bibitem{Leeks-2007} P. J. Leek, J. M. Fink, A. Blais, R. Bianchetti, M. G\"oppl, J. M. Gambetta, D. I. Schuster, L. Frunzio, R. J. Schoelkopf, A. Wallraff, Science, {\bf 318}, 1889 (2007).
\bibitem{Wallraff-2012} S. Berger, M. Pechal, S. Pugnetti, , A. A. Abdumalikov, Jr., L. Steffen, A. Fedorov, A. Wallraff, and S. Filipp, Phys. Rev. B {\bf 85}, 220502(R) (2012). 
\bibitem{Oberthaler-1999} C. L. Webb, R. M. Godun, G. S. Summy, M. K. Oberthaler, P. D. Featonby, C. J. Foot, and K. Burnett, Phys. Rev. A {\bf 60}, R1783.
\bibitem{Mottonen-2008} M. M\"ott\"onen, J. J. Vartiainen, J. P. Pekola, Phys. Rev. Lett. {\bf 100}, 177201 (2008).
\bibitem{Solinas-2010} P. Solinas, M. M\"ott\"onen, J. Salmilehto, and J. P. Pekola, Phys. Rev. B {\bf 82}, 134517 (2010).
\bibitem{Pekola-2010} J. P. Pekola, V. Brosco, M. M\"ott\"onen, P. Solinas, A. Shnirman, Phys. Rev. Lett. {\bf 105}, 030401 (2010).
\bibitem{Breuer} H.-P. Breuer, F. Petruccione, \textit{ The Theory of Open Quantum Systems}, (Oxford University, Oxford, 2002). 
\bibitem{Kamleitner-2011} I. Kamleitner and  A. Shnirman, Phys. Rev. B {\bf 84}, 235140 (2011).
\bibitem{Salmilehto-2012} J. Salmilehto, P. Solinas, M. M\"ot\"onen, Phys. Rev. A {\bf 85}, 032110 (2012).
\bibitem{levitov-1996} L. S. Levitov \textit{et al.}, J. Math. Phys. (N.Y.) {\bf 37}, 4845 (1996).
\bibitem{Belzig-Nazarov-2001} W. Belzig, Yu. V. Nazarov, Phys. Rev. Lett. {\bf 87}, 197006 (2001).
\bibitem{Niskanen-2003} A. O. Niskanen, J. P. Pekola, H. Sepp\"a, Phys. Rev. Lett. {\bf 91}, 177003 (2003).
\bibitem{mottonen-2006} M. M\"ott\"onen, J. P. Pekola, J. J. Vartiainen, V. Brosco, F. W. J. Hekking, Phys. Rev. B {\bf 73}, 214523 (2006).
\bibitem{Grifoni-1998} M. Grifoni, P. H\"anggi, Phys. Rep. {\bf 304} 229 (1998). 
\bibitem{Berry-1987} M. V. Berry, Proc. R. Soc. A {\bf 414}, 31-46 (1987).
\bibitem{footnote} Note  an error of factor $1/2$ in the temperature of Ref. \onlinecite{Kamleitner-2011}




\end{thebibliography}
\end{document}